\documentclass[a4paper]{JAC2003}

\usepackage{graphicx}
\usepackage{booktabs}
\usepackage{placeins}

\begin{document}
\title{Measurement of the dynamic response of the CERN DC spark system and preliminary estimates of the breakdown turn-on time}

\author{N. Shipman, S. Calatroni, R. M. Jones, W. Wuensch.}

\maketitle

\begin{abstract}
  The new High Repetition Rate (HRR) CERN DC Spark System has been used to investigate the current and voltage time structure of a breakdown.  Simulations indicate that vacuum breakdowns develop on ns timescales or even less.  An experimental benchmark for this timescale is critical for comparison to simulations.  The fast rise time of breakdown may provide some explanation of the particularly high gradients achieved by low group velocity, and narrow bandwidth, accelerating structures such as the T18 and T24.  Voltage and current measurements made with the previous system indicated that the transient responses measured were dominated by the inherent capacitances and inductances of the DC spark system itself.  The bandwidth limitations of the HRR system are far less severe allowing rise times of approximately 12ns to be measured.
\end{abstract}

\section{Introduction}

Many accelerator projects as well as other technologies and research areas involving large electromagnetic fields are beginning to run up against limits in performance imposed by breakdown.  In CLIC the rate of breakdown is a critical issue as it results in a loss of acceleration and a transverse kick to the beam which can cause a loss in luminosity.

Results of the RF tests indicate that low group velocity, and consequently narrow bandwidth structures are able to sustain much higher surface fields than high group velocity, large bandwidth, structures \cite{1a}\cite{group_v}. This dependency is captured by the high power limits $P/\lambda C$ and $S_c$ presented in \cite{group_v}. Reference \cite{group_v} also suggests a physical model to explain the origin of these limits and further study has led to the idea that the process which governs the turn on time is the instantaneous power flow available to feed the breakdown during its onset. In other words a high group velocity structure could more quickly replenish local energy density absorbed by a growing breakdown leading to faster turn on times.

Field emission currents at single emission sites are of the order of pA while the current during breakdown can be hundreds and even thousands of amps and as shown the transition between these two regimes, the turn on time, can be extremely fast.  An accurate measure of the rise time of breakdowns under various conditions is an essential step in understanding whether the transient response of RF systems to the breakdown currents determine breakdown limits.

\section{The CERN DC spark system}

The first CERN DC Spark Systems have been used for conducting quick and cost effective breakdown tests on different materials and to develop a better understanding of breakdown in general. The systems consist of an anode and cathode - the sample under test (copper in this paper) - in a point-plane geometry in ultra high vacuum.\cite{Old_System}

The previous DC spark system at CERN stored the energy needed for a breakdown on a capacitor and used a mechanical relay as the switching mechanism with which to apply the high voltage to the sample under test.  This method was limited in two respects: firstly the maximum repetition rate achievable was only around 0.5Hz, thus it would require several weeks of testing to reach a breakdown rate of $10^{-7}$ per pulse, the region of interest for CLIC; and secondly the frequency response of the circuit was complicated at high frequencies.

Recently a new High Repetition Rate (HRR) system has been designed in which coax and matched impedances are used throughout.  The energy for the breakdown is stored on a 200m long coaxial cable known as a Pulse Forming Line (PFL).  The mechanical relay has also been replaced by a solid state MOSFET switch which allows the system to operate at up to 1kHz and with a much faster switching time.  A detailed description of the HRR system is given in \cite{Rudi}.

\begin{figure*}[tb]
    \centering
    \includegraphics*[width=168mm]{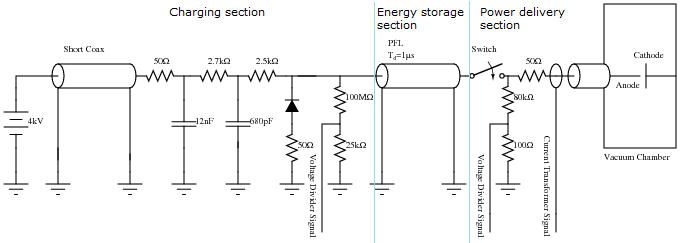}
    \caption{Circuit diagram of the new high-rep-rate CERN DC spark system.
Use of coax, matched impedances and the fast switching time of the MOSFET switch allow much better insight into the turn on time of the breakdown.}
    \label{Circuit}
\end{figure*}

As indicated in Fig.\ref{Circuit} the circuit consists of three parts: the charging section; the energy storage section; and the power delivery section.  The charging section contains a power supply whose voltage can be set between zero and $12$kV and which can deliver a maximum current of $10$mA.  This section charges the energy storage section to the selected voltage.  The energy storage section contains the PFL, this is where the energy which will be dissipated during a breakdown is stored.  The total capacitance of the PFL is $2$nF, thus it can store up to $\sim 1.4$J at $12$kV, the PFL can be totally discharged in $2\mu$s corresponding to a power of $0.7$MW.  The power delivery section consists of the MOSFET switch a $50\Omega$ matching resistor and the spark gap.

Whilst the switch is open the PFL charges up to the selected voltage, always 4kV for measurements in this paper.  When the switch closes this voltage is all dropped across the gap which normally has a resistance far greater than the $50\Omega$ resistor.  If the sample then breaks down the resistance of the gap becomes very low and the PFL discharges through the $50\Omega$ resistor and the spark gap to ground.  Once the PFL has fully discharged and the breakdown has extinguished the switch opens and the PFL is charged up again ready for the next pulse.

\section{Simulations}

\begin{figure}[htb]
	\hspace{-5mm}
   \includegraphics*[width=90mm]{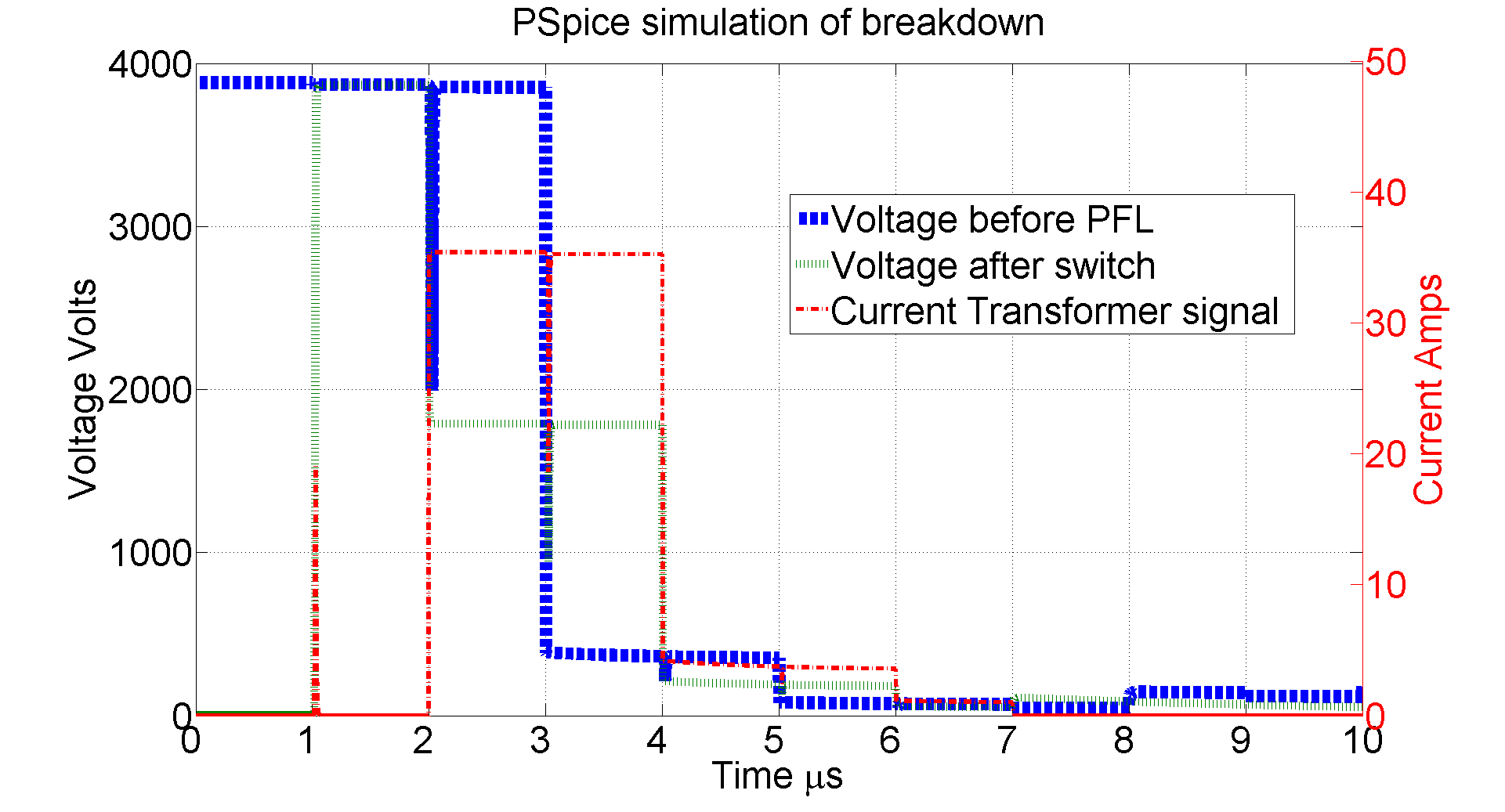}
   \caption{Breakdown simulation}
   \label{BD_sim}
\end{figure}

In order to ensure that the HRR circuit was well understood we modelled a simplified version in PSpice\cite{PSpice}.  The results of this simulation are shown in Fig.\ref{BD_sim}. Simplifying assumptions include the characteristics of the MOSFET switch the lack of various stray reactances and the resistance of the PFL which was set to the measured DC value.  The spark gap was also modelled as a switch which was closed for breakdown and open otherwise.

Three signals with which to compare simulation with measurement were chosen, these were the ``voltage before the PFL'', the ``voltage after the switch'' and the ``current transformer signal''.  The current measured by the Current Transformer (CT) is that which flows through the spark gap (and that required to charge the stray capacitance).  The voltage after the switch is a sum of the voltage drop across the spark gap and the $50\Omega$ resistor.

Fig.\ref{BD_sim} shows the simulated values of these signals during a breakdown.  The first current spike is that required to charge stray capacitances between the switch and spark gap.  Signals due to the reflection of this spike from the ends of the PFL can be seen on all the measured signals at multiples of $1\mu$s.  The current quickly drops to zero again until the sample breaks down and a larger current flows for $2\mu$s, the time it takes to discharge the PFL.

The voltage after the switch rises to $4$kV when the switch is closed before dropping to approximately half that value during the breakdown as the resistance of the gap becomes very small and some of the voltage is dropped across the PFL and most of the rest across the $50\Omega$ resistor.

\begin{figure}[htb]
	\hspace{-5mm}
   \includegraphics*[width=90mm]{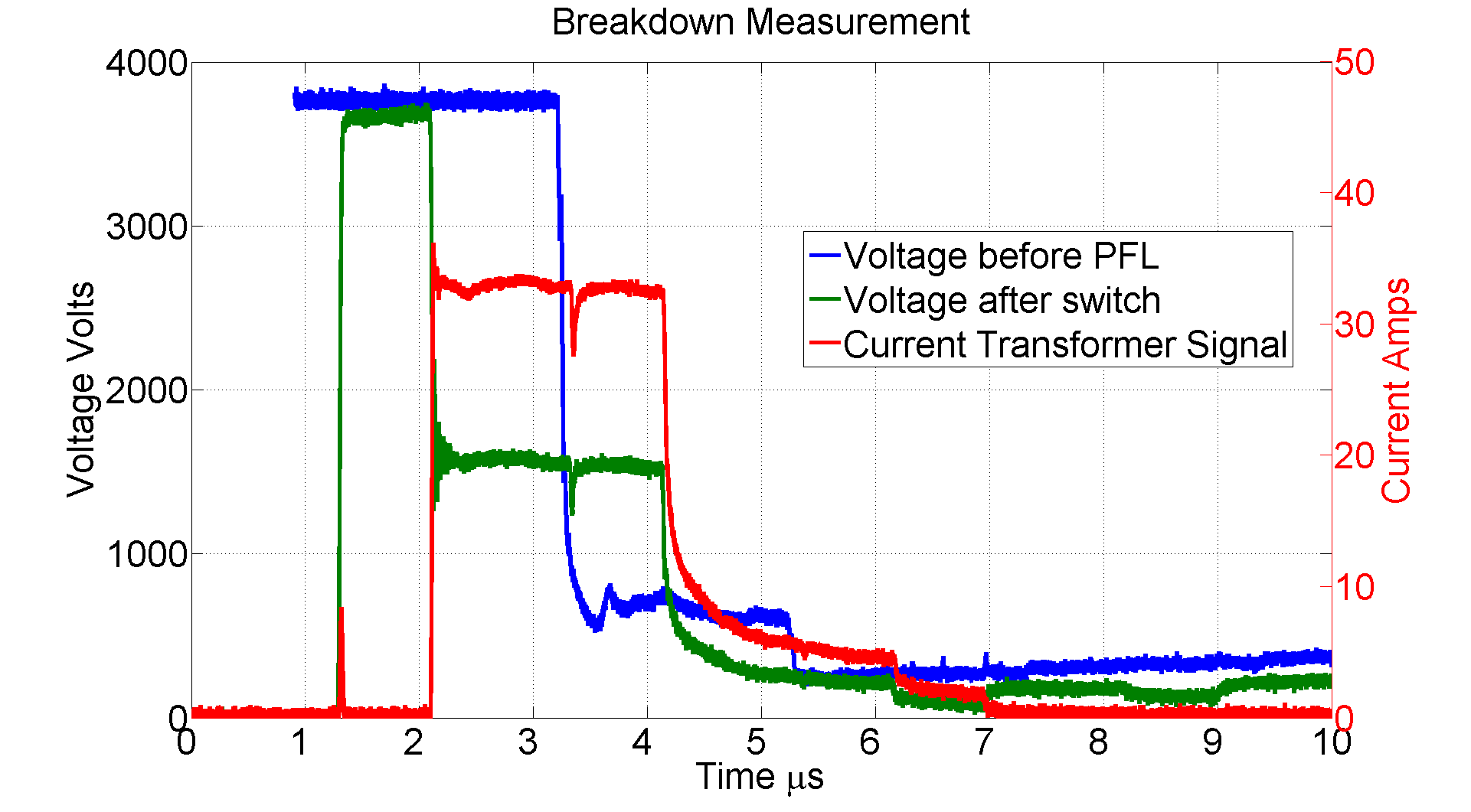}
   \caption{Breakdown measurement.}
   \label{BD_Meas}
\end{figure}

The voltage before the PFL stays at $4$kV for $\sim 1\mu$s after the breakdown starts, the time it takes for the signal to propagate down the line before dropping to nearly zero.

\section{Circuit Behaviour}

Fig.\ref{BD_Meas} shows measurements of the circuit behaviour made for comparison to the simulation. Despite the simple model used the agreement between the simulations and measurements was generally good with the same basic features.  All the transitions in the real system were generally less sharp than in the simulation and the amplitudes of the reflections were larger and of shorter duration, this is likely due to stray impedances and other bandwidth limitations.  We would also expect the measured voltage after the switch to drop by half following the breakdown but it actually drops to less than this. This is likely due to a slight miss match in impedances somewhere and will be investigated further.

\section{Breakdown turn on time and burning voltage measurements}

\begin{figure}[htb]
	\hspace{-5mm}
   \includegraphics*[width=90mm]{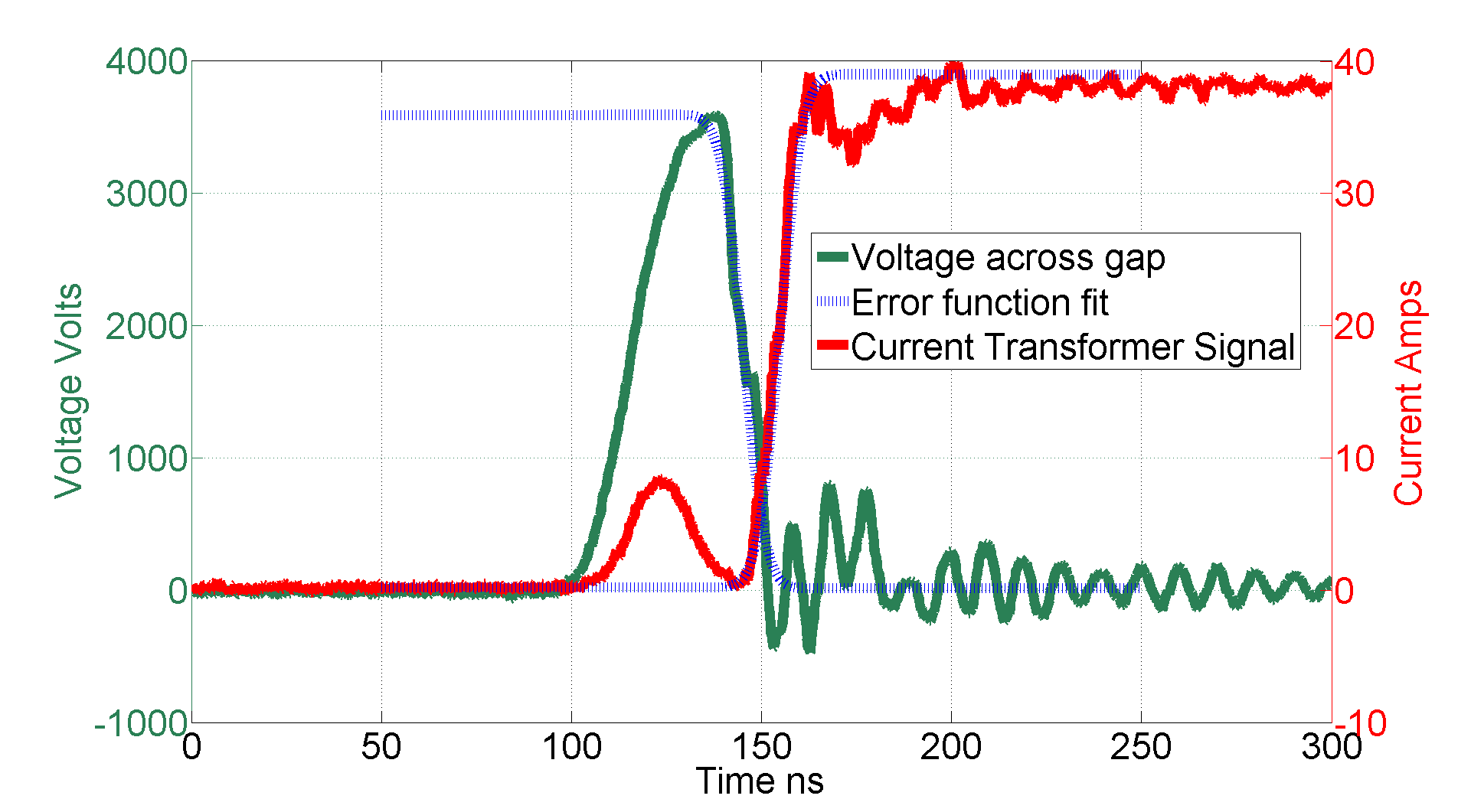}
   \caption{Rise time and current voltage measurements across gap.}
   \label{tran}
\end{figure}

Fig.\ref{tran} shows the measured voltage over the gap (the other side of the $50\Omega$ resistor than was measured previously) and the current  measured with the CT during the initiation of a breakdown. The voltage measurement was made with a Tektronix P6015A voltage probe with a $75$MHz bandwidth and the current measuremnts with a Bergoz CT-B0.5 current transformer with a bandwidth of $200$MHz.  Both signals were read into a LeCroy wavepro 7100 digital osciloscope with a bandwidth of $1$GHz.  Using these bandwidths and the well known formula $BW*T_r = 0.35$\cite{Rise_Time} the smallest voltage and current rise times measureable were estimated to be $5$ns and $2$ns respectively.

To facilitate clear comparisons between this measurement and others, an error function style fit (eqn.\ref{eqn}) was performed on both signals and we calculate the $10\%$ to $90\%$ rise (or fall times)\cite{erf}.
\begin{eqnarray}\label{eqn}
V \propto (1-erf(\alpha\tau))\\
I \propto (1+erf(\alpha\tau))
\end{eqnarray}

We measure $12$ns for both the rise time of the current and fall time of the voltage.  Whilst these measured rise times do not appear to be limited by the bandwidth of the measurement techniques, it is not clear if they are measures of the intrinsic turn on time of the breakdown or, as hypothesised in the introduction, are governed by the bandwidth of the system; which Fourier transforms of the oscillations after the transient suggest is $\sim 0.1$GHz.

The CLIC accelerating structures have a similar bandwidth and the falling edges of transmitted RF signals are also similar typically a few $10$s of ns.  The previous CERN DC Spark system also had a much lower bandwidth and the measured current rise time there was larger still at $~150$ns\cite{Jan_Thesis}.  Whilst a confident quantitative prediction of the rise time is still just beyond the scope of current particle in cell breakdown simulation efforts, rise times in the order of a few ns often appear, even shorter than this latest measurement\cite{Helga_Thesis}.

With this measurement we have clearly shown that the intrinsic turn on time of a breakdown, if there is such a thing, is less than $12$ns, we would have been easily able to measure it with our system if it had been longer.  In the future we will reduce stray impedances in order to increase the bandwidth and see if the turn on time is further reduced.  Dedicated measurements and analysis of the falling edge of transmitted power due to breakdown in RF structures is keenly awaited so the results may be compared.  Comparison of the results together with the bandwidth of the systems will be able to show whether the turn on time in RF strucutres is limited by the bandwidth of the structures, which could help explain how such low breakdown rates have been achieved in low group velocity structures.

After the breakdown transit has occurred the gap will develop a constant voltage, known as the burning voltage, which is normally around $20$V for clean copper\cite{burning_v}.  By averaging the measured voltage from $250$ns to $400$ns we obtain a burning voltage of $11$V, quite consistent to the value above.
In the future efforts will be made to reduce the ringing as well as improve the dynamic range so more accurate measurements can be made.

\vspace{-15pt}
\section{ACKNOWLEDGMENT}
We would like to thank Mike Barnes, Rudi Soares and Jan Kovermann for their patient help, expert advice and useful discussion as well as Paul Garritty for all the time he spent in the workshop making various pieces which made these measurements possible.

\end{document}